\documentclass{jps-cp}
\usepackage{txfonts} 
\usepackage{bm}
\allowdisplaybreaks[4]

\title{Accessing Polarized Fragmentation Functions at the Unpolarized EIC and BELLE Experiments}

\author{Shu-Yi \textsc{Wei}$^{1}$}

\inst{$^{1}$Institute of Frontier and Interdisciplinary Science, Key Laboratory of Particle Physics and Particle Irradiation (MOE), Shandong University, Qingdao, Shandong 266237, China}

\email{shuyi@sdu.edu.cn}

\abst{We briefly report our recent progress on the study of the polarized fragmentation functions of $\Lambda$ hyperon in unpolarized semi-inclusive deep-inelastic scatterings and electron-positron annihilations at low energies. In particular, we present a simple but practical method on how to measure the azimuthal-angle-dependent longitudinal polarization and the transverse polarization inside the production plane and bridge these observables to the corresponding structure functions and fragmentation functions at the leading twist. Our work diversifies the high energy reactions that can probe the polarized fragmentation functions.}

\kword{SIDIS, $e^+e^-$ annihilation, Polarization, Fragmentation Function.}

\begin{document}

\maketitle

\section{Introduction}

Fragmentation functions (FFs) are non-perturbative quantities that offer insight into the hadronization process of a high energy parton. The polarized FFs, with an additional dimension, can deliver more information about non-perturbative physics. However, the polarized FFs are in general less studied due to the scarcity of experimental data. 

The situation has been improved a bit recently after Belle collaboration published the first measurement of $\Lambda$-hyperon transverse polarization in $e^+e^-$ annihilation \cite{Belle:2018ttu}. The measurement is conducted as follows. In the center-of-mass frame of the incoming electron and positron, two almost back-to-back jets are produced. From one jet, we measure a $\Lambda$-hyperon and from the other, we measure a light hadron. The transverse polarization of $\Lambda$ is measured along the normal direction of this production plane, which is defined by the momenta of $\Lambda$ and the light hadron. The readers may have already figured out that this experiment is designed to probe $D_{1T}^\perp (z, p_T)$, if they are familiar with the physical interpretations of leading twist FFs. Several parameterizations \cite{DAlesio:2020wjq,Callos:2020qtu,Chen:2021hdn} and model calculations \cite{Li:2020oto} have been promptly carried out.

To probe the flavor dependence of $D_{1T}^\perp$, Belle experiment measured the transverse polarization of $\Lambda$ with different associated light hadrons. The philosophy of this measurement is that the light hadron can put a flavor tag on the parent quark. To our surprise, the difference between the polarization in $\Lambda+\pi^+$ and $\Lambda+\pi^-$ is distinct. The same goes for $\Lambda+K^+$ and $\Lambda + K^-$ processes as well. The D'Alesio-Murgia-Zaccheddu (DMZ) \cite{DAlesio:2020wjq} and Callos-Kang-Terry (CKT) \cite{Callos:2020qtu} parameterizations introduce a significant isospin symmetry violation to describe the Belle data. Moreover, a model calculation \cite{Li:2020oto} suggests it cannot describe the data without the sea parton contribution under the constraint of the $SU(6)$ spin-flavor symmetry.

The isospin symmetry is a robust symmetry in QCD. A strong violation is not expected in the polarized FFs where the strong interaction dominates. Therefore, we explore the possibility to describe Belle data with an isospin symmetric parameterization for the $D_{1T}^\perp$ FF. We also propose to ultimately test the isospin symmetry in the unpolarized semi-inclusive deep inelastic scatterings (SIDIS) at large Bjorken $x$. With the ability to perform both $ep$ and $eA$ scatterings, the future EIC can easily accomplish this task. 

Furthermore, we also propose other polarization observables in the unpolarized SIDIS and low energy $e^+e^-$ annihilations, albeit it might be counter-intuitive. These observables are azimuthal angle dependent and therefore vanish in the $2\pi$ phase space average. The simple method proposed in our work makes it possible to access the polarized FFs at the future electron-ion collider (EIC) or the current Belle experiment. Our work diversifies high energy processes that can probe the polarized FFs and therefore is a complementary study in the field.

This proceeding is organized as follows. In Sec.~2, we discuss the isospin symmetry of FFs and present an isospin symmetric parameterization that can well describe the Belle data. In Sec.~3, we briefly demonstrate the idea of accessing the polarized FFs in the unpolarized SIDIS and low energy $e^+e^-$ annihilation. In Sec.~4, we briefly discuss the possibility to probe the longitudinal spin transfer $G_{1L}$ and the longitudinal-to-transverse spin transfer $G_{1T}^\perp$ at Belle. We make a summary in Sec.~5.

\section{Isospin symmetry}

The Belle data \cite{Belle:2018ttu} shows a distinct difference between the transverse polarizations of $\Lambda$ in $\Lambda+\pi^+$ and $\Lambda+\pi^-$ processes. A naive picture to interpret this data is that $\Lambda+\pi^+$ favors the $d\to \Lambda$ and $\bar d\to \pi^+$ channel, since this is the only combination that both $\Lambda$ and $\pi^+$ are produced from the favored (valence) parton. Using the same argument, we may find that $\Lambda+\pi^-$ production favors the $u\to \Lambda$ and $\bar u \to \pi^-$ channel. Therefore, from this naive picture, we can get the impression that $D_{1T u}^{\perp \Lambda} \neq D_{1T d}^{\perp \Lambda} \neq D_{1T s}^{\perp \Lambda}$. Following this perspective, two isospin-asymmetric parameterizations \cite{DAlesio:2020wjq,Callos:2020qtu} have been proposed recently and describe the Belle data quite well.

However, this naive picture is a good approximation only at large $z_\Lambda$ and large $z_h$. At small $z_\Lambda$ and $z_h$, the flavor components are far more complicated than what the aforementioned picture suggests. Therefore, we investigate the flavor components in each process and conclude that the difference between the transverse polarizations in $\Lambda+\pi^+$ and $\Lambda+\pi^-$ processes can be easily explained by taking into account contributions from unfavored (sea) partons.

Using the Trento convention \cite{Bacchetta:2004jz} for $D_{1T}^\perp$, we identify the transverse polarization of $\Lambda$ as
\begin{align}
{\cal P}_N (z_\Lambda, z_h, |\bm{P}_{\Lambda\perp}|) = \frac{{\cal I} [\omega_1 D_1^h D_{1T}^{\perp\Lambda}]}{{\cal I}[D_1^h D_1^\Lambda]},
\end{align} 
where $z_\Lambda$ is the longitudinal momentum fraction carried by $\Lambda$, $z_h$ is that carried by the light hadron and $\bm{P}_{\Lambda\perp}$ is transverse momentum of $\Lambda$ with respect to the light hadron momentum. Here, we have used the shorthand notation ${\cal I} [\omega D^h D^\Lambda] \equiv \sum_q e_q^2\int d^2\bm{p}_T d^2\bm{p}_{hT} \delta^2 (\frac{z_\Lambda}{z_h} \bm{p}_{hT} + \bm{p}_T - \bm{P}_{\Lambda\perp}) \omega D^h (z_h, \bm{p}_{hT}) D^{\Lambda} (z_\Lambda, \bm{p}_{T})$ with $\bm{p}_{hT}$ and $\bm{p}_{T}$ being the transverse momentum of light hadron and $\Lambda$ with respect to the parent partons. $\omega_1$ is given by
$\omega_1 \equiv - \frac{\hat P_{\Lambda\perp} \cdot p_T}{z_\Lambda M_\Lambda}$,
where $\hat P_{\Lambda\perp}$ is the unit four-vector along the direction of ${P}_{\Lambda\perp}$.

\begin{figure}[htb]
\centering
\includegraphics[width=0.3\textwidth]{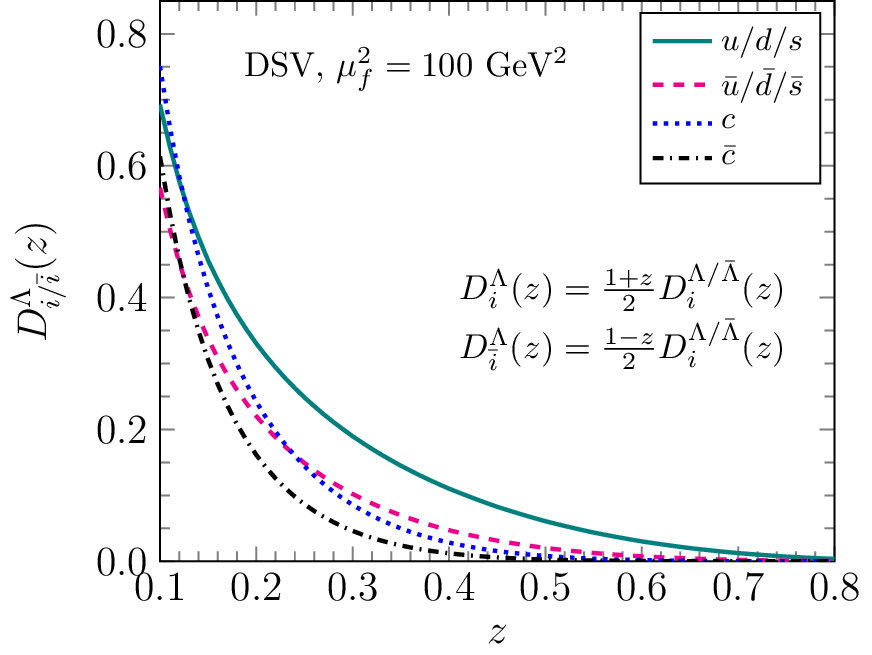}
\includegraphics[width=0.3\textwidth]{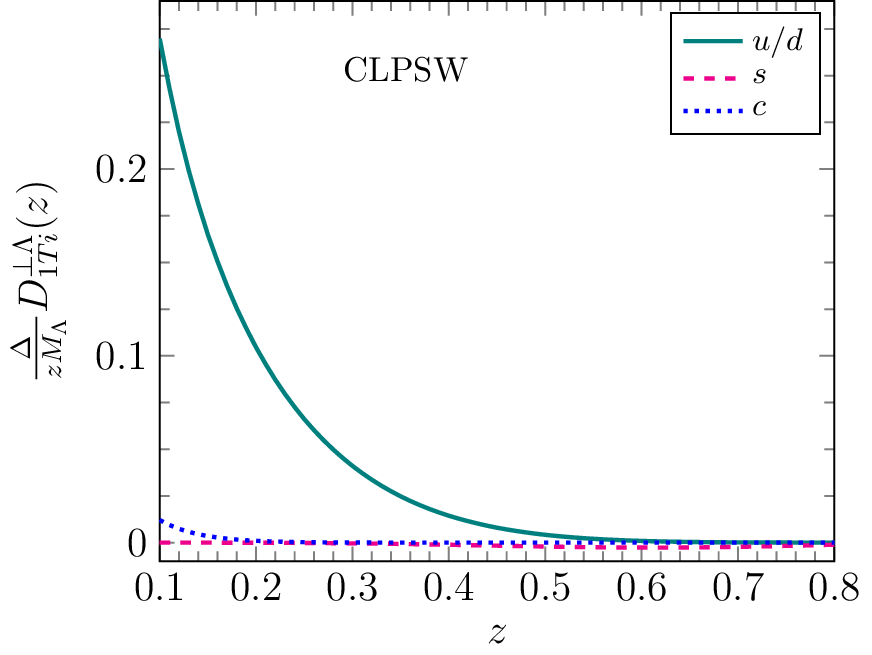}
\includegraphics[width=0.3\textwidth]{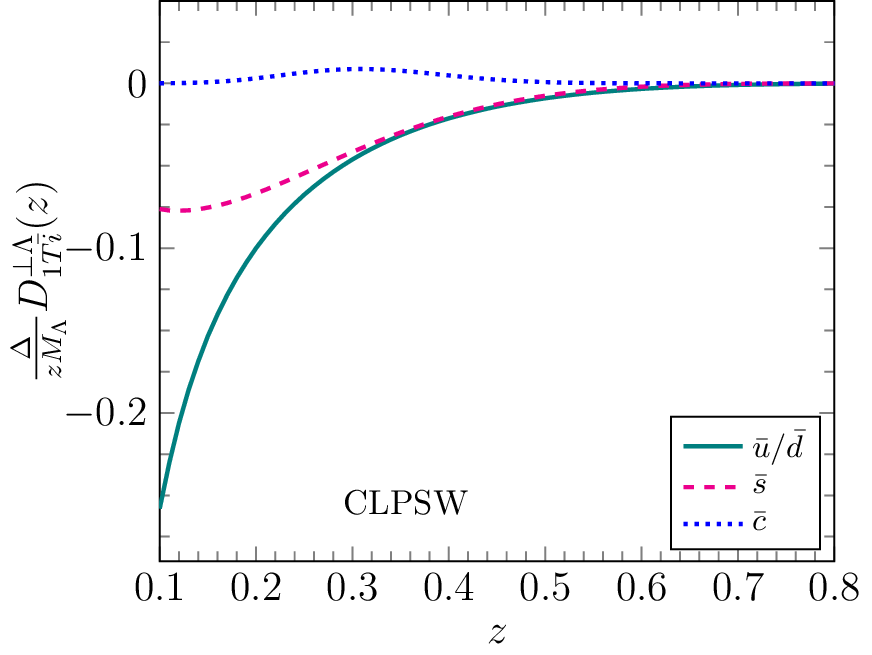}
\caption{The unpolarized and polarized FFs of $\Lambda$ hyperon. The unpolarized FFs of $\Lambda$ is related to the $\Lambda/\bar\Lambda$ FFs provided by DSV parameterization \cite{deFlorian:1997zj} via the prescription labeled on the plot. 
}
\label{fig:ffs}
\end{figure}

To proceed, we assume that the $z$ dependence and the $p_T$ dependence in FFs can factorize, while the $p_T$ dependence part takes the Gaussian ansatz. Therefore, we have 
\begin{align}
D_{1Ti}^{\perp\Lambda} (z,\bm{p}_{T}) = D_{1Ti}^{\perp\Lambda} (z) \frac{1}{\pi \Delta^2} \exp\left[-\frac{p_T^2}{\Delta^2}\right],
\end{align}
where $\Delta$ is the Gaussian width which is assumed to be flavor-independent for simplicity. This Gaussian assumption is a common practice in parameterizing the TMD PDFs/FFs at the initial scale. The evolution of these TMD functions is governed by the well-known Collins-Soper equation which is given in Refs.~\cite{Collins:1981uk}. Therefore, the $p_T$ distributions of TMD PDFs/FFs usually deviate from the simple Gaussian function at large scale. Nonetheless, we decide to focus on the $p_T$-integrated observables since the number of data points is limited. The exact form of the $p_T$ distribution does not matter any more for the $p_T$-integrated observables. It only contributes to a constant factor which can be absorbed into the overall normalization. For the unpolarized FFs of $\Lambda$ and light hadron, we employ the DSV \cite{deFlorian:1997zj} and DEHSS \cite{deFlorian:2014xna} parameterizations. With these setups, we extract the parameterization of $D_{1Ti}^{\perp\Lambda} (z)$ under the constraint of isospin symmetry by fitting the Belle data and show the polarized and unpolarized FFs of $\Lambda$ in Fig.~\ref{fig:ffs}.

This isospin symmetric parameterization can also describe the Belle data well. Therefore, we demonstrate that the Belle experimental data does not automatically translate into isospin symmetry violation in $D_{1T}^\perp$. Furthermore, we propose a clean test on this point at the future EIC experiment \cite{Chen:2021zrr}. At large $x$, the dominant contribution to nucleon PDF comes from $u$ and $d$ quarks. At the future EIC experiment, we can switch the target from a proton to a large nucleus and therefore can tune the effective $u$ and $d$ quark PDFs. In light of this, we can easily conclude that there will be almost no difference between the transverse polarizations in $ep$ and $eA$ scatterings at large-$x$. However, if the isospin symmetry is indeed strongly violated as the DMZ and CKT parameterizations suggested, we shall observe an evident difference. The above argument solely relies on the dominance of $u$ and $d$ quark PDFs at large-$x$. The uncertainties in the polarized FFs and unpolarized FFs can affect the exact value of the predicted transverse polarization, but will not undermine this qualitative conclusion.

\section{Longitudinal polarization in unpolarized SIDIS}

This concept might be counter-intuitive to some readers. At first sight, it seems to violate parity conservation to measure the longitudinal polarization in the unpolarized SIDIS. However, the physics behind this is quite simple. Although the parent nucleon is unpolarized, the partons inside it can still be transversely polarized thanks to the Boer-Mulders function. This transverse polarization can further propagate to the azimuthal asymmetries and longitudinal and transverse polarizations of final state hadrons through chiral-odd FFs. It enriches the observables that can be studied in the most simple process. In this proceeding, we use longitudinal polarization as an example to demonstrate how to measure these observables. A more general discussion is given in Ref.~\cite{Chen:2021zrr}.

From the kinematic analysis, we can easily find the longitudinal polarization of the $\Lambda$ hyperons produced in the unpolarized SIDIS takes the following form,
\begin{align}
{\cal P}_L (x,y,z, \bm{P}_{\Lambda\perp})
=
\frac{C(y) \sin(\phi) F_{UL}^{\sin \phi} + B(y) \sin(2\phi) F_{UL}^{\sin 2\phi}}
{A(y) F_{UU}^{T} + B(y) F_{UU}^{L} + C(y) \cos(\phi) F_{UU}^{\cos\phi} + B(y) \cos(2\phi) F_{UU}^{\cos2\phi}},
\end{align}
where $F$'s are scalar structure functions. 

Clearly, if we only select events where $\Lambda$ hyperons are produced in the first and second quadrants, the $\sin(2\phi)$ structure in the numerator does not contribute. The $\cos(\phi)$ and $\cos(2\phi)$ contributions also disappear. Therefore, we can probe this $F_{UL}^{\sin \phi}$ structure function. Similarly, if we only select events where $\Lambda$ hyperons are produced in the first and third quadrants, the $\sin (\phi)$, $\cos(\phi)$ and $\cos(2\phi)$ structures vanish. We can then probe the $F_{UL}^{\sin2\phi}$ structure function. At the leading order and leading twist approximation, $F_{UL}^{\sin \phi} = 0$ and  $F_{UL}^{\sin 2\phi}$ is related to the convolution of Boer-Mulder function and $H_{1L}^\perp$ FF. Assuming that the magnitude of $H_{1L}^\perp$ is roughly the same with that of $D_{1T}^\perp$ and using the Boer-Mulders function extracted from the Drell-Yan process, we estimate the longitudinal polarization can be as large as a few percent and should be measurable at the future EIC in light of its high luminosity.

Another approach is to measure the $\sin(\phi)$ or $\sin(2\phi)$ weighted polarization. It is straightforward to show these two approaches are equivalent. Moreover, by measuring the azimuthal-angle-dependent transverse polarizations, we can study the chiral-odd $H_{1T}$ and $H_{1T}^\perp$ FFs.

Following the same strategy, we can also access polarized chiral-odd FFs in low energy $e^+e^-$ annihilation by measuring two back-to-back hadrons. In this case, the Collins function of the reference light hadron plays the role to provide information on the transverse polarization of parent quarks. Utilizing this method, we open new opportunities for probing polarized chiral-odd FFs. 


\section{Dihadron polarization correlation in $e^+e^-$ annihilation}

Similar ideas can further extend to other observables as well. For instance, we can study the longitudinal spin transfer $G_{1L}$ and the longitudinal-to-transverse spin transfer $G_{1T}^\perp$ at Belle as well. 

The longitudinal spin transfer $G_{1L}$ describes the probability of producing a longitudinally polarized hadron from a longitudinally polarized parton. It has been studied at LEP by measuring the longitudinal polarization of the produced $\Lambda/\bar\Lambda$ \cite{ALEPH:1996oew,OPAL:1997oem}. The hard interaction is dominated by weak interaction at LEP and therefore the quarks are longitudinally polarized. However, at low energy $e^+e^-$ colliders, the hard interaction is dominated by electromagnetic interaction. It is very difficult, if not impossible, to probe the longitudinal spin transfer at Belle. 

However, the longitudinal polarization of the quark and that of the antiquark produced in the same hard interaction are correlated. We then define the polarization correlation of two back-to-back $\Lambda/\bar\Lambda$-hyperons, ${\cal PC}_L (z_1,z_2)$, as the possibility of same-sign polarization minus that of the opposite-sign polarization. At the leading order and leading twist approximation, the dihadron polarization correlation is related to $G_{1L}^{q\to\Lambda/\bar\Lambda} (z_1, p_{T1}) \otimes G_{1L}^{\bar q\to \Lambda/\bar\Lambda} (z_2, p_{T2})$. Furthermore, it is straightforward to measure this observable in experiments as well. Using $\theta_{i}^*$ to denote the angle between the longitudinal direction and the momentum of $p/\bar p$ in the $\Lambda/\bar\Lambda$-rest frame, we find 
\begin{align}
\frac{dN}{d\cos\theta_1^* d\cos\theta_2^*} = 1 + \alpha {\cal P}_L (z_1) \cos\theta_1^* + \alpha {\cal P}_L (z_2) \cos\theta_2^*+ \alpha^2 {\cal PC}_L (z_1,z_2) \cos\theta_1^* \cos\theta_2^*,
\end{align}
where $\alpha$ is the decay parameter, ${\cal P}_L (z_i)$ is the longitudinal polarization of $\Lambda/\bar\Lambda$ with momentum fraction $z_i$ and ${\cal PC}_L (z_1,z_2)$ is the aforementioned dihadron polarization correlation. At low energy, ${\cal P}_L (z_i) \approx 0$ and ${\cal PC}_L (z_1,z_2) \neq 0$. Thus, we can probe the longitudinal spin transfer by measuring $\langle \cos\theta_1^* \cos\theta_2^* \rangle$ at low energy $e^+e^-$ annihilation. Furthermore, we can study the longitudinal-to-transverse spin transfer, $G_{1T}^\perp$, by measuring the longitudinal-transverse or transverse-transverse polarization correlation of two back-to-back $\Lambda/\bar\Lambda$-hyperons. 

A similar idea has also been proposed in Ref.~\cite{DAlesio:2021dcx} recently. We refer interested readers to Ref.~\cite{DAlesio:2021dcx} for more details.

\section{Summary}

In this proceeding, we discuss the capacity to eventually test isospin symmetry of $D_{1T}^{\perp\Lambda}$ at the future EIC experiment. We also present how to measure the azimuthal-angle-dependent longitudinal polarization and transverse polarizations in unpolarized SIDIS and low energy $e^+e^-$ annihilation. With these novel observables, we can access chiral-odd FFs in the simple processes, which have been carried out for decades.

\end{document}